\documentclass[12pt,preprint]{aastex}
\usepackage{color}
\usepackage{natbib}

\begin{document}
\shorttitle{Saturn-mass planet in the HZ of HIP 57050}
\shortauthors{Haghighipour et al.}

\title{The Lick-Carnegie Exoplanet Survey:
A Saturn--Mass Planet in the Habitable Zone 
of the Nearby M4V Star HIP 57050}
\author{Nader Haghighipour \altaffilmark{1}, 
Steven S. Vogt \altaffilmark{2},
R. Paul Butler \altaffilmark{3}, Eugenio J. Rivera \altaffilmark{2} , 
Greg Laughlin\altaffilmark{2}, Stefano Meschiari \altaffilmark{2}, 
Gregory W. Henry \altaffilmark{4}}

\altaffiltext{1}{Institute for Astronomy and NASA 
Astrobiology Institute,
University of Hawaii-Manoa, Honolulu, HI 96822}
\altaffiltext{2}{UCO/Lick Observatory, Department 
of Astronomy and Astrophysics,
 University of California at Santa Cruz, Santa Cruz, CA 95064}
\altaffiltext{3}{Department of Terrestrial Magnetism, 
Carnegie Institute of Washington, Washington, DC 20015}
\altaffiltext{4}{Center of Excellence in Information Systems,
Tennessee State University, Nashville, TN 37209}

\begin{abstract}
Precision radial velocities from Keck/HIRES reveal a Saturn-mass 
planet orbiting the nearby M4V star HIP 57050. The planet has a 
minimum mass of $M \sin i \sim 0.3 M_{\rm J}$, an orbital period of 
41.4 days, 
and an orbital eccentricity of 0.31. V-band photometry reveals a 
clear stellar rotation signature of the host star with a period of 
98 days, well separated from the period of the radial velocity variations 
and reinforcing a Keplerian origin for the observed velocity variations. 
The orbital period of this planet corresponds to an orbit in the habitable 
zone of HIP 57050, with an 
expected planetary temperature of $\sim$230 K. The star has a metallicity of 
[Fe/H] = $0.32\pm0.06$ dex, of order twice solar and among the highest 
metallicity stars in the immediate solar neighborhood. This newly 
discovered planet provides further support that the well-known 
planet-metallicity correlation for F, G, and K stars also extends down 
into the M-dwarf regime. The {\it a priori} geometric probability for 
transits of this planet is only about 1\%. However, the expected eclipse 
depth is $\sim 7\%$, 
considerably larger than that yet observed for any transiting planet. 
Though long on the odds, such a transit is worth pursuing as it would 
allow for high quality studies of the atmosphere via transmission 
spectroscopy with HST. At the expected planetary effective temperature, 
the atmosphere may contain water clouds.
\end{abstract}


\section{Introduction}

Due to their low masses and surface temperatures, M dwarfs present the
most promising targets for searching for terrestrial-mass and
potentially habitable planets. As the least massive stars, these objects
experience the greatest reflex accelerations in response to an orbiting 
planet. This advantage was first realized with the detection of 
a Neptune-mass extrasolar planet around the star GJ 436 \citep{Butler04}, 
and the first super-Earth around the star GL 876 \citep{Rivera05}.
The low surface temperatures of M dwarfs place their (liquid water) 
habitable zones at conservative distances of approximately 0.1 AU to 0.2 AU. 
These distances correspond to orbital periods of 20 to 50 
days, implying another advantage of M dwarfs
as potential targets for detecting habitable planets relatively quickly.

As precision Doppler surveys are 
optimally sensitive to small orbits, it is not surprising that 
terrestrial-mass planets around M dwarfs, in particular those in the 
habitable zone, have been the subject of research for more than a decade 
\citep{Joshi97,Segura05,Boss06,Scalo07,Grenfell07,Tarter07}. 
During the past few years, such research resulted in the detection of
17 extrasolar planets around 12 M dwarfs\footnote{We refer the reader 
to exoplanet.eu for more 
details.}. Slightly more than half of these planets are Neptune-mass or 
smaller, 
consistent with the fact that M dwarfs have smaller circumstellar 
disks, and experience has shown that they are less frequently accompanied by 
readily detectable planets, and/or their planets are less massive 
compared to those of G stars.

While the majority of the currently known extrasolar planets have been 
detected around nearby F,G, and K stars, more than 70\% of the nearest 
stars are M dwarfs. For the past decade, we have had a sample of $\sim$300 
nearby quiet stars under precision radial velocity survey (P.I. 
Butler's NASA M-dwarf Exoplanet Survey) with the Keck telescope and its 
HIRES spectrometer. Here, we present 9.9 years of precision radial 
velocities for the nearby M4 dwarf HIP 57050 and report the detection
of the exoplanet they imply.

\section{HIP 57050}

HIP 57050 (LHS 2443, GJ 1148) is an M4 dwarf \citep{Reid04}
with a $V$ magnitude of 
$11.881 \pm 0.004$ and color $B-V=1.60$ 
\citep{Perry97,Kharchenko01}. The distance of this star,
as obtained from its Hipparcos parallax ($90.66 \pm 3.03$ mas) 
\citep{Perry97}, is $11.0 \pm 0.4$ pc, making this star one
of the nearest M dwarfs. 

The SIMBAD listed J, H, and K magnitudes of HIP 57050
are 7.608, 7.069, and 6.822, respectively \citep{Cutri03}. 
Given its distance, the corresponding absolute J, H, and K 
magnitudes of this M dwarf are 7.401, 6.862, and 6.615.
The empirical mass-luminosity relation of \citet{Henry93} 
can be used to estimate the mass of this star.
From the values of the
J, H, and K absolute magnitudes of HIP 57050, the mass
of this M dwarf is approximately $0.34 \pm 0.03 \, {M_{\odot}}$.

As shown by \citet{Morales08}, HIP 57050 has an effective
temperature of ${T_{\rm eff}}=3190$ K, and an 
absolute bolometric magnitude of 
${M_{\rm bol}}= 9.32$. Assuming the bolometric magnitude of 
the Sun to be ${M_{{\rm bol}, \, \odot}}=4.75$, we compute  
the luminosity of HIP 57050 to be $L=0.01486 {L_{\odot}}$.
Comparison of the luminosity and effective temperature of
HIP 57050 with those of the Sun suggests a 
radius of $0.4 {R_{\odot}}$. We measure a chromospheric 
activity index of $\log {R'_{\rm HK}}=-5.31$, implying an expected 
jitter of 1.9 m\,s$^{-1}$. Table 1 summarizes the stellar parameters
of HIP 57050.

\section{Radial Velocity Observations}

A total of 37 precision radial velocities of HIP 57050 
were obtained with the 
HIRES spectrometer \citep{vog94} at the Keck observatory. 
Doppler shifts were measured by placing an Iodine 
absorption cell just ahead of the spectrometer slit in the converging
beam from the telescope \citep{but96}. 
This gaseous Iodine absorption cell superimposes 
a rich forest of Iodine lines on the stellar spectrum, providing a 
wavelength calibration and proxy for the point spread function (PSF) 
of the spectrometer. The Iodine cell is sealed and temperature-controlled 
to 50.0$\pm$0.1 $^{\circ}$C so that the column density of Iodine remains 
constant. We operate the HIRES spectrometer at a spectral resolving power 
of $R \sim 70,000$ and wavelength range of 3700\,--\,8000 \AA.
Only the region 5000\,--\,6200 \AA\, (with Iodine lines) was
used in the present Doppler analysis. 
The Iodine region is divided into $\sim$700 
chunks of 2 \AA\,  each. Each chunk produces an independent measure of 
the wavelength, PSF, and Doppler shift. The final measured velocity 
is the weighted mean of the velocities of the individual chunks.

Observations were carried out for over 9.9 years from February 2000 till 
January 2010. Table 2 and Figure 1 show the individual observations.
The median internal uncertainty for our observations is
2.8 m\,s$^{-1}$, and the peak-to-peak velocity variation is 
95.1 m\,s$^{-1}$.  The velocity scatter about the mean RV in our measurements 
is 24.5 m\,s$^{-1}$. Figure 2 shows the periodogram of the RVs and
the power spectral window (PSW) of our sampling.
In the top panel of this figure, the plotted power is proportional to
the relative improvement in the fit quality for the best Keplerian 
fit found at that period versus
a constant velocity model. The bottom panel of Figure 2 shows the
spectral window or power due to the sampling times \citep{Deeming75}. This
spectral window indicates spurious power that can be introduced into the
data from the sampling times alone.

Several methods have been presented to define and normalize the power  
as in the top panel of Figure 2. For instance, \citet{GB87} consider
an error-weighted Lomb-Scargle periodogram, and renormalize the power,
relative to the noise, at some interesting peak using
\begin{equation}
p_0 = \frac{1}{4}\,N\, {x_0^2}\,{\sigma_0^{-2}}.
\end{equation}
\noindent
In this equation, $x_0$ is the fitted RV half-amplitude implied by 
the peak, and $\sigma_0$ is the RV scatter in the data prior to fitting 
out the implied signal. \citet{Cumming04}, on the other hand, 
defined the power at each trial period as 
\begin{equation}
p_0 = A \,\, \frac{\chi_{\rm constant}^2-\chi_{\rm model}^2}
{\chi_{\rm constant}^2}\,,
\end{equation}
\noindent
where $\chi_{\rm constant}^2$ is the reduced $\chi^2$ for a constant RV model
(the mean of the RVs), $\chi_{\rm model}^2$ is the reduced $\chi^2$ for 
a model, which could, for example, 
be a simple sinusoid or a Keplerian orbit, 
and $A$ is a normalization factor that depends on the number of 
observations and the number of fitted parameters. 
To estimate the false alarm probability (FAP) of a given peak
with either of the above-mentioned definitions, 
the knowledge of the number of independent frequencies in the data set
$(M)$ is required. Both \citet{GB87} and
\citet{Cumming04} give procedures to estimate $M$. Through experimentation, we
have developed guidelines that enable us to roughly relate the values of 
$M$ obtained from equations (1) and (2) together.
These rough relations save substantial computing time when we model Keplerian
orbits at all trial periods as we do in the top panel of Figure 2.  
At each trial period, we fit a Keplerian orbit with various initial
values for the eccentricity, longitude of periastron, and mean anomaly.  
The power in the top panel of Figure 2 corresponds to the best-fit 
Keplerian orbit of all these fits. The guidelines mentioned above lead us to
estimate the FAP for the strong (Keplerian) signal in Figure 2 to be
$<10^{-7}$. The horizontal lines in this figure represent,
from top to bottom, the 0.1\%, 1.0\%, and 10.0\% FAP levels,
respectively. Additionally, the FTEST probability for our best one-planet 
fit is $2.2\times10^{-8}$.

\section{Keplerian Modeling of the Radial Velocity Observations}

Our fitting was carried out with the publicly available 
{\it Systemic Console} \citep{systemic}.
The velocity zero point is arbitrary and was allowed to float as
part of the fitting process. A Keplerian orbital 
fit to the radial velocity data of HIP 57050 (Figure 3) points
to the existence of a planet with a minimum mass of $0.3 M_{\rm J}$
and an orbital eccentricity of 0.31. Table 3 shows the orbital elements
of this planet. As shown by the periodogram of our data (Figure 2), 
the planet's orbital period is 41.4 days corresponding to a
semimajor axis of $\sim$0.16 AU. 

We also examined the possibility of additional companions in the system.
Table 3 also lists the parameters for a fit to the RV data for
HIP 57050 consisting of one planet plus a linear trend. In comparing the two
fits, the FTEST probability indicates that the trend is not significant.
However, examination of subsets of the data indicates that the trend is a
plausible realization for all the cases examined since all fits to the
examined subsets result in comparable slopes to the linear trend.  Also, the
addition of a linear trend has a significant effect on $\chi_{\nu}^2$ as well
as on the periodogram of the residuals.  Figure 4 shows the periodograms
of the residuals for the fits presented in Table 3.
The power in these periodograms is based on fitting  
circular orbits at each trial period.
There is a peak with FAP $<0.001$ near 16 days in the periodogram of the
residuals of the one-planet fit.  The FTEST probability for this second
companion is also very small at 0.00043 which suggests that this is a
viable solution.  However, the relatively small number of observations, the
plausibility of a trend in the RVs and its effect on the periodogram of the
residuals, and the uncertain status of the stability of a two-planet
fit with the second planet at 16 days cast some doubt as to what is the
correct best fit for the current RV set.  Additionally, if we use the method of
\citet{GB87} to obtain the periodogram for the one-planet residuals, 
we find the
FAP of the most prominent peak, which is also near 16 days, to be $>0.1$.
More data will be required to verify or refute either solution which 
would indicate
the presence of a second companion.

If we assume that the inner boundary of the habitable zone (HZ) 
of the Sun is at
0.95 AU \citep{Kasting93}, and its outer boundary is at a distance between
1.37 AU and 2.4 AU, depending on the chosen atmospheric circulation model
\citep{Forget97,Mischna00}, then by direct comparison, the inner 
boundary of the HZ of HIP 57050 would be at a distance of $\sim$0.115 AU, 
and its outer boundary would be between
0.163 AU and 0.293 AU. From Table 3, the perihelion and aphelion distances of
HIP 57050 b are at 0.112 AU and 0.215 AU respectively, suggesting that this 
planet spends the majority of its orbital motion in the HZ of its host star. 
Although the planet makes small excursions outside the HZ,
due to the response time of the atmosphere-ocean system 
\citep{Williams02,Jones06}, 
and the effect of ${\rm CO}_2$ cloud circulations 
\citep{Selsis07,Forget97,Mischna00}, the times of these 
excursions are small compared to the time that is necessary for a significant 
change in the temperature of the planet to occur.
In other words, the planet could hardly be more squarely in the HZ
and will most likely maintain its habitable status 
even when its orbit is temporarily outside of this region.

 \section{Photometric Observations}

We acquired Johnson $V$ photometry of HIP 57050 during the 2006--2007 and 
the 2007--2008 observing seasons with an automated 0.35 m Schmidt-Cassegrain 
telescope and an SBIG ST-1001E CCD camera. This Tennessee State University 
telescope was temporarily mounted on the roof of Vanderbilt University's 
Dyer Observatory in Nashville, Tennessee.

We computed differential magnitudes of HIP 57050 for each epoch of 
observation from ten consecutive CCD images with exposure times in the 
range of 10--20 seconds. Our differential magnitudes represent the 
difference in brightness between HIP 57050 and the mean of five constant 
comparison stars in the same field of view, averaged over the ten CCD 
frames at each epoch. Outliers from each group of ten images were removed 
based on a $3\sigma$ test. If three or more outliers were filtered from 
any group of ten CCD frames (usually the result of non-photometric 
conditions), the entire group was discarded. The final standard deviations 
of the nightly means ranged from 0.001--0.005 mag, depending on the quality 
of the night. One or two mean differential magnitudes were acquired each 
clear night. Our final data set consists of 548 observations spanning 563 
nights.  

Our goal with the photometric observations of HIP 57050 was to look for 
signs of activity and, if present, to find the star's rotation period from 
the rotational modulation of features on the star's photosphere 
\citep[see, e.g.,][]{Henry95}. These observations help to determine if 
the radial velocity variations are caused by intrinsic stellar activity 
\citep{Queloz01} or by stellar reflex motion caused by the presence of 
an orbiting companion. We discarded the first 82 and the last 48 days of 
photometric measurements so that the remaining portion of the light curve 
exhibits reasonably coherent variability. We also discarded a few obvious 
outliers from the shortened light curve, which retains 314 measurements 
ranging over 433 days (see the top panel of Figure 5). Cyclic variability 
is easy to see. It is obvious from the top panel of Figure 5 that HIP 57050 
is varying in 
brightness over a range of a couple percent on a timescale of approximately 
100 days. A bootstrap analysis gives a rotation period of $98.1 \pm 0.6$ days. 
The solid line corresponds to the sum of a 98.1
days rotation period and a second component with a longer period for removing the
season-to-season drift. The missing  portion of the light curve 
is due to the lack of observation between the two observing seasons. 

The second panel of Figure 5 shows the power spectrum of the photometric
data. As shown here, a strong periodicity exists near 98 days which is
presumably due to spots on the star rotating at this rate.
The third panel shows the spectrum of residuals from our best-fit 
photometric period of $98 \pm 0.6$ days. The peak near 333 days is due to the
season-to-season drifts in brightness which can be attributed to the long-term 
changes
in the spot distribution. We have modeled this drift as the partial phase of
a second sinusoidal component of period 328 days.
The lesser peak near 45.5 days in the third panel is not significant and disappears
when the 328 days component is fitted out, leaving no significant power
at or near the 41.4 days Keplerian period in modeling either the full or
shortened data sets.

The photometric observations are replotted in the bottom panel 
of Figure 5. Here the season-to-season baseline drift has been removed, 
and observations have been phased to the 98.1-day rotation period and a 
time of minimum computed from a least-squares sine fit. 
The sine fit also gives the 
peak-to-peak amplitude of 0.023 mag. The phase curve is slightly 
asymmetrical with the ascending branch shorter than the descending branch, 
as is often seen in the light curves of active stars 
\citep[see][]{Henry95}. That this period is clearly well separated 
from the radial velocity period argues strongly against stellar 
rotation as being the cause of the velocity variations and provides 
additional support for a planetary origin for the observed velocity 
variations.

\section{Discussion}

In the quest for potentially habitable planets, the nearest stars are
of special importance. They have accurate distances and precisely
determined stellar parameters, and are the only stars for which
follow up by astrometry and direct imaging is possible.
Within the Sun's immediate neighborhood, M-dwarfs
constitute the majority of nearby stars. As such, these stars have the
special properties (distances, masses, and habitable zones) that
drive exoplanetary science, astrobiology, and the next generation
of interferometry and direct imaging missions. 
The habitable (liquid water) zones of nearby M-dwarfs are typically between 
0.1 AU 
and 0.2 AU which corresponds to orbits with periods of 20 to 50 days. 
Establishing (by direct detection) the prevalence and nature of low-mass 
planets, such as HIP 57050 b, in these orbits informs us greatly about 
the possibility for
potentially habitable planets (and/or moons) in the solar neighborhood.

A zeroth-order prediction of the core-accretion paradigm for giant 
planet formation is that the frequency of readily detectable giant 
planets should increase with both increasing stellar metallicity and 
with increasing stellar mass \citep{Laughlin04,Ida05}. 
During the past decade, both of these trends have been 
established observationally \citep[see, e.g.][for a 
discussion of the metallicity trend and Johnson et al. 2009 for a 
discussion of the mass trend]{Fischer05}.
Until recently, however, there appeared to be little evidence for the 
strong expected planet-metallicity correlation among the handful of 
M-dwarf stars that are known to harbor giant planets. 
Attempts to determine accurate metallicities of M-dwarfs
have largely been stymied by ambiguity in the continuum levels of their
heavily line-blanketed spectra and by the profusion of molecular
features in their spectra. Conventional 
estimates for the metallicities of 7 of the currently known 
planet-hosting M dwarfs as given by \citet{Bailey09}, and a comparison 
between these estimates and those of \citet{Schiavon97} and \citet{Bean06}, 
suggest a spread of metallicity among these M-dwarfs 
(four are metal poor, one has high metallicity, and the metallicities 
of the remaining two are solar).

One would naively expect that a low-mass disk will need all the help 
it can get in order to build giant planet cores before the gas is 
gone. If anything, the planet-metallicity correlation should be 
strongest among the M-dwarfs. If observations show that the 
planet-metallicity correlation breaks down for M-dwarfs, then one is 
naturally led to speculate that the infrequent giant planets in a systems 
like Gliese 876 might be the outcome of gravitational instability 
\citep[e.g.][]{Boss00} rather than core accretion.

\citet{Bonfils05} pioneered a new approach to the determination of 
M-dwarf metallicities. The long evolutionary time scales for M-dwarfs 
imply that age-related $L$ and $T_{\rm eff}$ changes should be minimal once 
a low-mass star has landed on the zero-age main sequence. M-dwarf 
positions on the color-magnitude diagram, therefore, should be 
parameterized only by mass and metallicity, opening the possibility of 
a metallicity determination based on ${\rm M}_{\rm K}$ 
and V-K alone. \citet{Bonfils05} 
developed such a calibration by assuming that M-dwarf binary 
companions to F, G, and K stars share the readily determined 
metallicities of their primaries. \citet{Johnson09} have recently 
provided an update to the \citet{Bonfils05} calibration. The 
\citet{Johnson09} calibration indicates that the planet-bearing 
M-dwarfs do appear to be systematically metal-rich, suggesting that 
there is no breakdown of the planet-metallicity correlation as one 
progresses into the red dwarf regime.

HIP 57050 b appears to offer further support for the emerging M-dwarf 
planet-metallicity correlation. Using 
HIP 57050's values, $V=11.88$, $K=6.822$, and $d=11.03\,{\rm pc}$, 
the \citet{Johnson09} calibration yields 
$[{\rm Fe/H}]=0.32\pm0.06\,{\rm dex}$, 
indicating that HIP 57050 has a metallicity of order twice solar, which 
places it among the highest metallicity stars in the immediate solar 
neighborhood.

The {\it a priori} geometric transit probability for HIP 57050 b is $\sim
1\% $. The small size of the primary star and the planet's unfavorable
orbital alignment ($\omega=238^{\circ}$) conspire to diminish the odds
that transits can be observed. An analysis of our photometry data also
shows no signs of a transit. However, because the orbital elements can change
if a second planet emerges, it is premature to conclude at this point
that transits do not occur. The eclipse depth in this system is expected 
to be $\Delta F/F \sim 7\%$, which is considerably larger than that yet 
observed for any transiting planet. Such a large depth makes this system
suitable for small-telescope observers to check. We
therefore suggest that small-telescope observers carry out photometric
monitoring of HIP 57050 during the predicted transit windows centered on
HJD 2455201.400239.
The large planet-to-star ratio would allow for detailed study of the  
atmosphere via transmission spectroscopy with HST. The expected  
planetary effective temperature, $T_{\rm eff}\sim230\,{\rm K}$,  
suggests that the atmosphere may contain water clouds.

It is interesting to speculate about the possible presence of a habitable 
moon around HIP 57050 b. By analogy with our own solar system, whose gas 
giants all have dozens of moons, one might expect HIP 57050 b to also harbor 
such moons. In our solar system, $\sim0.02\%$ of the masses of the gas giants 
are assigned to their satellites. This would translate to a satellite with 
$\sim2\%$ of Earth's mass (similar to Titan) orbiting HIP 57050 b. 
While it is not out of the question that HIP 57050 b could
harbor a moon, and that moon would thus be in the liquid water habitable
zone of the parent star, an object with only 1/5th of the mass of Mars
in the liquid water habitable zone, from various standpoints is probably
not a particularly good prospect for habitability. In any case, direct
detection of such a moon would be extremely challenging.

We conclude this study by noting that
the Doppler radial velocity method continues to be the most productive  
and cost-effective way to find those extrasolar planets that impart the  
greatest scientific returns 
\citep{Butler04,Rivera05,Lovis06,Udry07_HARPS,Mayor08,vogt10,Rivera10}. 
During the past several years, the threshold $M\sin(i)$ for  
radial velocity planets has rapidly approached the 1$M_{\oplus}$ regime. 
Radial velocity surveys, furthermore, have led to the  
discovery of all but one of the most readily characterizable  
transiting planets, and the rapidly growing 
catalog of Doppler-detected planets has  
been instrumental in providing our best current view of the nearby  
planetary population\footnote {see the planet lists and correlation  
diagrams at exoplanet.eu}. The future looks bright!

\acknowledgments
NH acknowledges support from the NASA Astrobiology Institute under 
Cooperative Agreement NNA04CC08A at the Institute for Astronomy, University
of Hawaii, and NASA EXOB grant NNX09AN05G. SSV 
gratefully acknowledges support from NSF grants AST-0307493 and AST-0908870,
and from the NASA Keck PI program. RPB gratefully acknowledges support 
from NASA OSS Grant NNX07AR40G, the NASA Keck PI program, and from 
the Carnegie Institution of Washington. GL acknowledges support 
from NSF AST-0449986. GWH acknowledges support from NASA, NSF, 
Tennessee State University, and the state of Tennessee
through its Centers of Excellence program. We also gratefully 
acknowledge the major contributions over the past decade of fellow 
members of our previous California-Carnegie Exoplanet team: 
Geoff Marcy, Jason Wright, Debra Fischer, and Katie Peek in helping 
to obtain some of the radial velocities presented in this paper. 
We are also thankful to the anonymous referee for a careful review of our
paper and his/her suggestions that have improved our manuscript.
The work herein is based on observations obtained at the W. M. Keck 
Observatory, which is operated jointly by the University of California 
and the California Institute of Technology, and we thank the UH-Keck, 
UC-Keck and NASA-Keck Time Assignment Committees for their support. 
We also wish to extend our special thanks to those of Hawaiian 
ancestry on whose sacred mountain of Mauna Kea we are privileged 
to be guests. Without their generous hospitality, the Keck 
observations presented herein would not have been possible. 
This research has made use of the SIMBAD database, operated 
at CDS, Strasbourg, France.

\clearpage
\begin{figure}
\center
\includegraphics[angle=-90, scale=0.5]{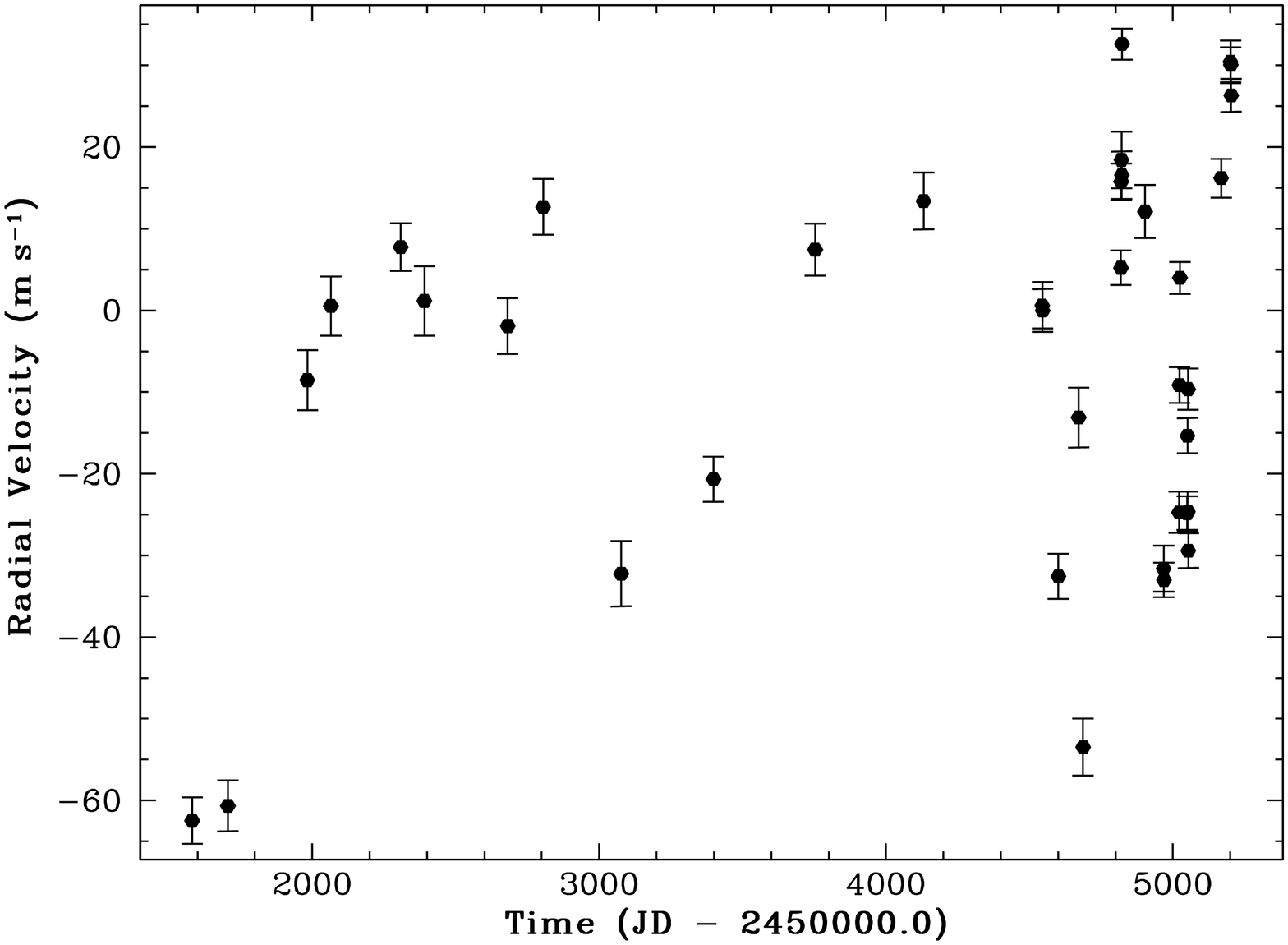}
\caption{Relative radial velocities for HIP 57050 obtained 
with the HIRES spectrometer on the Keck I telescope. The zero point 
is arbitrary and set to the mean of all the velocities.}
\end{figure}

\clearpage
\begin{figure}
\center
\includegraphics[angle=-90, scale=0.4]{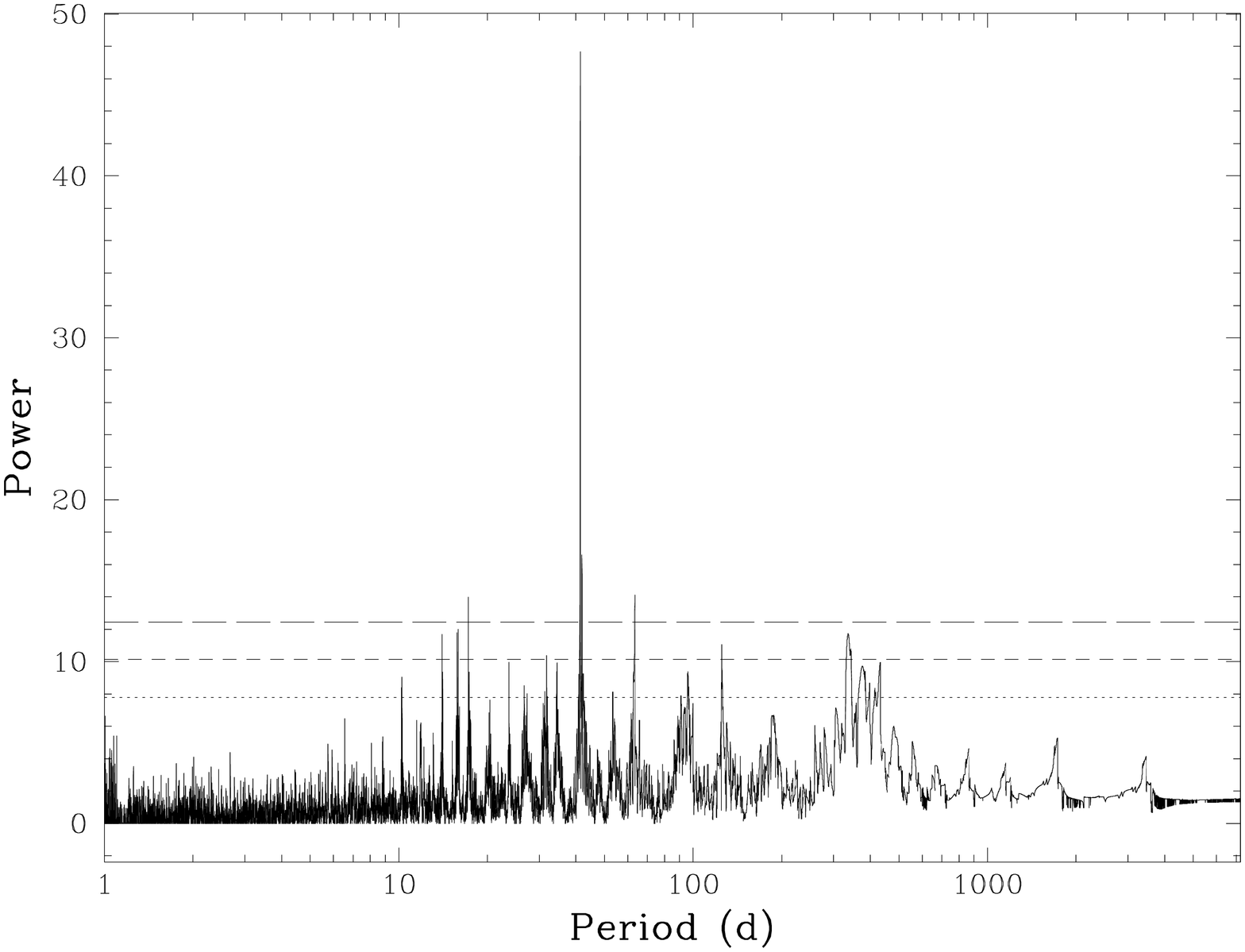}
\includegraphics[angle=-90, scale=0.4]{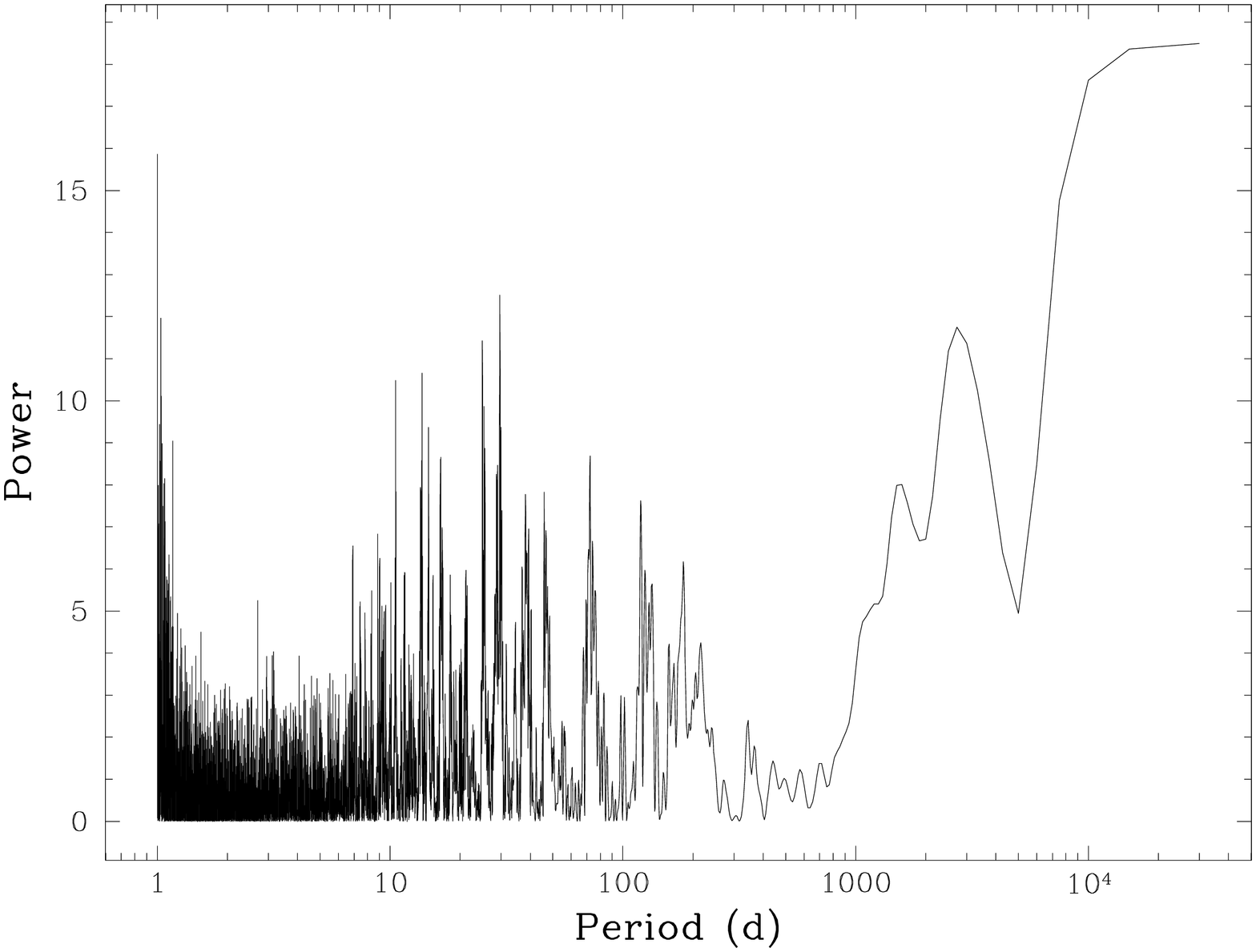}
\caption{
{\it Top panel} - Keplerian periodogram of the radial velocity data set for
HIP 57050. The power plotted at each sampled period is proportional to
the relative improvement (drop in $\chi_{\nu}^2$) in the fit
quality for the best Keplerian found at that period 
versus a constant velocity model.
The horizontal lines in this and all similar
figures indicate (top to bottom) False Alarm Probability (FAP) levels of
0.1\%, 1\%, and 10.0\% respectively. The dominant peak in the top panel
corresponds to the best-fit Keplerian orbit.
{\it Bottom panel} - Power spectral window (periodogram of the times of
observation).}
\end{figure}

\clearpage
\begin{figure}
\center
\includegraphics[angle=-90, scale=0.5]{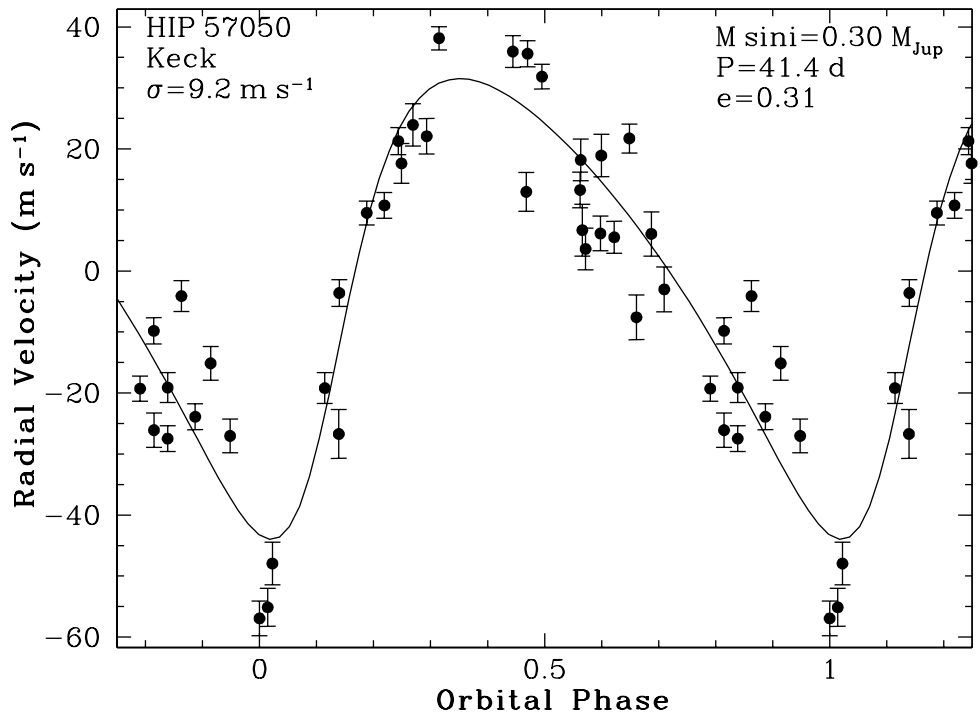}
\caption{Best one-planet Keplerian fit to the phased Keck-HIRES 
relative radial velocities of HIP 57050.}
\end{figure}

\clearpage
\begin{figure}
\center
\includegraphics[angle=-90, scale=0.5]{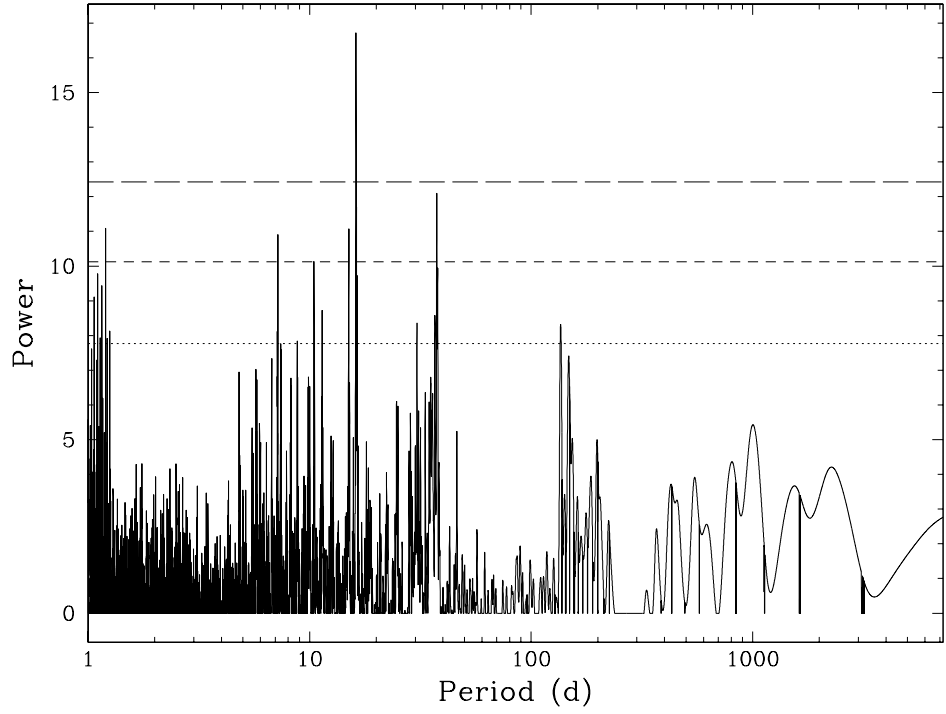}
\vskip 20pt
\includegraphics[angle=-90, scale=0.5]{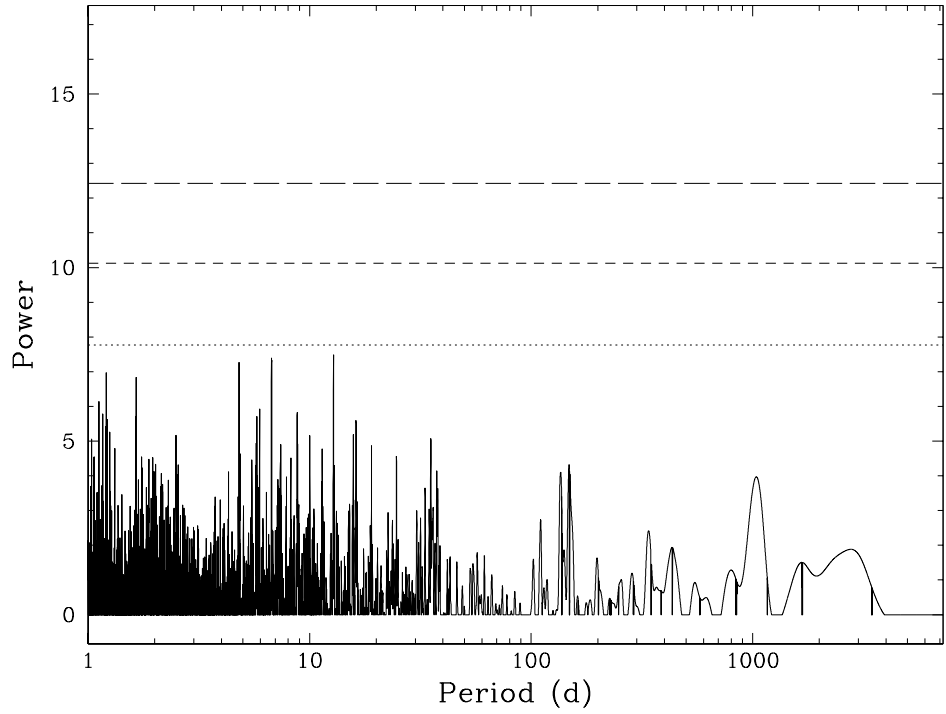}
\caption{Top: Circular periodogram of residuals from a 1-planet fit.
Bottom: Circular periodogram of residuals from a 1-planet fit+trend fit.}
\end{figure}

\clearpage
\begin{figure}[p]
\epsscale{0.8}
\plotone{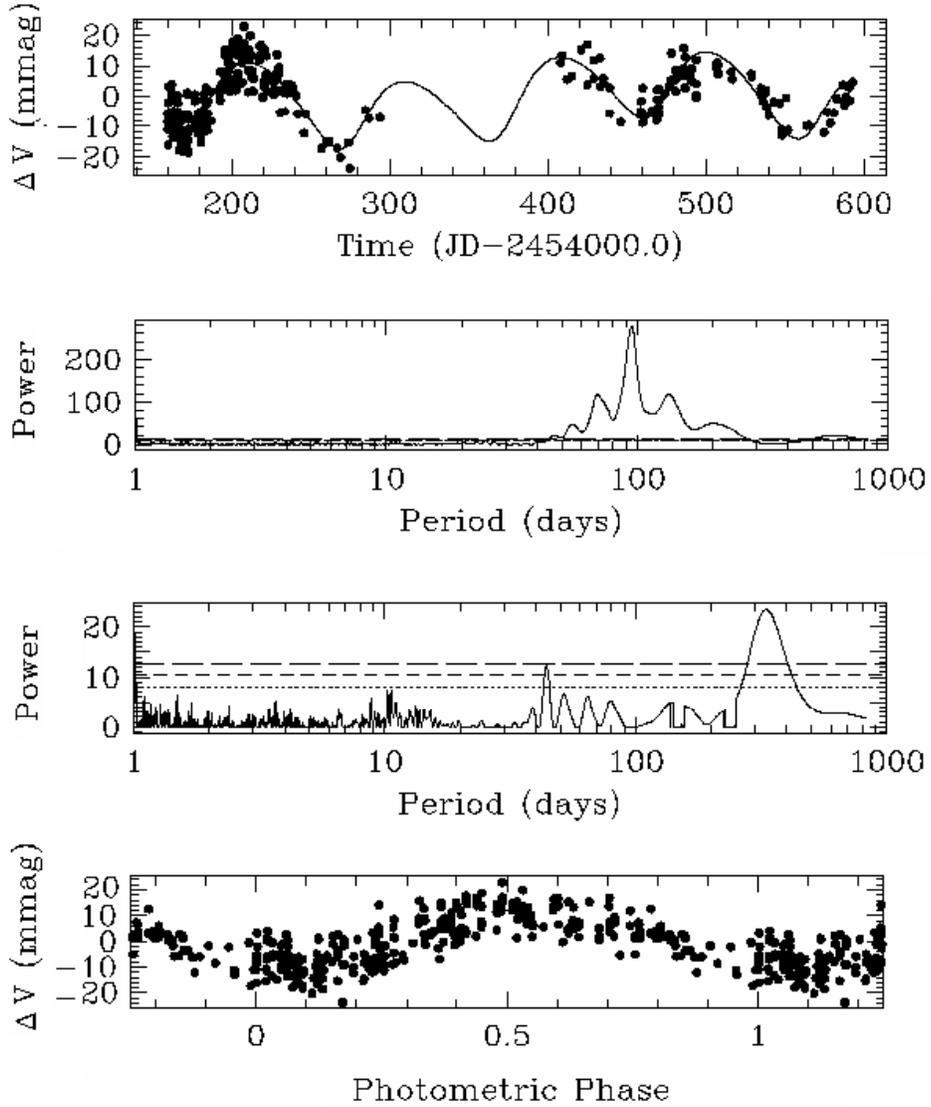}
\caption{
{\it Top panel} : Johnson V-band photometric observations of HIP 57050
from the 2006-07 and 2007-08 observing seasons acquired with a 0.35 m
imaging telescope. The fitted line is the sum of a 98.1 days rotation period
plus a second longer period component to track the seasonal drift.
{\it Second panel down} : Power spectrum of the photometric data revealing
strong periodicity, presumably due to spots on a star rotating at a
period near 98 days.
{\it Third panel down} : Power spectrum of the residuals from a best-fit
photometric period of $98.1 \pm 0.6$ days. The dominant peak near 333 days
reflects season-to-season drifts in brightness due to long-term changes
in the spot distribution. This drift is modeled here as the partial
phase of a second sinusoidal component of period 328 days.
{\it Bottom panel} : V-band observations from the top panel with
season-to-season baseline drift removed, and phased with the 98.1 day
photometric period. The peak-to-peak amplitude of the brightness
variation is 0.023 mag.}
\end{figure}

\clearpage
\begin{deluxetable}{lll}
\tabletypesize{\scriptsize}
\tablecolumns{3}
\tablewidth{0pt}
\tablecaption{Stellar Parameters for HIP 57050}
\tablehead{
Parameter & Value & Reference
}
\startdata
\label{stellarparams}
Spectral Type         & M4                & \citet{Reid04} \\
Mass ($M_{\odot}$)    & 0.34$\pm$0.03     & This work \\
Radius ($R_{\odot}$)  &0.4                & This work \\
Luminosity ($L_{\odot}$) & 0.01486        & This work \\
Distance (pc)         & 11.0$\pm$0.4      & \citet{Perry97} \\
$B-V$                 & 1.60              & \citep{Perry97,Kharchenko01}\\
$V$ Mag.              & 11.881$\pm$0.004  & \citep{Perry97,Kharchenko01}\\
$J$ Mag.              & 7.608             & \citep{Cutri03}\\
$H$ Mag.              & 7.069             & \citep{Cutri03}\\
$K$ Mag.              & 6.822             & \citep{Cutri03}\\   
$\log{R'_{\rm HK}}$   & -5.31             & This work \\
$P_{\rm rot}$ (days)  & 98                & This work \\
$T_{\rm eff}$ (K)     & 3190              & \citet{Morales08} \\
$M_{{\rm bol},\,\odot}$ & 9.32              & \citet{Morales08} \\
$\log{g}$             & 4.67              & This work \\
\enddata
\end{deluxetable}

\clearpage
\begin{deluxetable}{lcc}
\tabletypesize{\scriptsize}
\tablecaption{Relative Radial Velocities for HIP 57050}
\tablewidth{270pt}
\tablehead{
JD (-2450000) \qquad\qquad\qquad & RV (m\,s$^{-1}$) 
\qquad\qquad & Uncertainty (m\,s$^{-1}$) \\
}
\startdata
1581.04559 &   -62.47 &  2.83\\
1705.82690 &   -60.65 &  3.12\\
1983.00875 &    -8.54 &  3.66\\
2064.86395 &     0.55 &  3.63\\
2308.07715 &     7.74 &  2.92\\
2391.03363 &     1.16 &  4.27\\
2681.05010 &    -1.92 &  3.41\\
2804.88465 &    12.67 &  3.42\\
3077.10434 &   -32.24 &  3.99\\
3398.97476 &   -20.67 &  2.78\\
3753.06771 &     7.44 &  3.18\\
4131.09206 &    13.39 &  3.47\\
4545.00223 &     0.62 &  2.84\\
4546.00720 &     0.00 &  2.62\\
4600.90598 &   -32.56 &  2.76\\
4671.81115 &   -13.12 &  3.66\\
4686.77023 &   -53.48 &  3.49\\
4819.09864 &     5.22 &  2.10\\
4820.10874 &    15.76 &  2.21\\
4821.17169 &    18.42 &  3.48\\
4822.16958 &    16.55 &  2.91\\
4823.07178 &    32.62 &  1.91\\
4903.13549 &    12.10 &  3.26\\
4967.95619 &   -31.62 &  2.80\\
4968.94631 &   -33.00 &  2.12\\
5021.75704 &   -24.72 &  2.54\\
5022.80694 &    -9.14 &  2.18\\
5024.80704 &     3.99 &  1.96\\
5049.74445 &   -24.81 &  2.05\\
5050.74198 &   -15.34 &  2.16\\
5051.74421 &   -24.64 &  2.44\\
5052.74328 &    -9.64 &  2.52\\
5053.74585 &   -29.42 &  2.12\\
5168.06193 &    16.20 &  2.37\\
5201.00003 &    30.44 &  2.61\\
5202.07289 &    30.08 &  2.12\\
5203.11589 &    26.33 &  2.02\\
\enddata
\label{velocities}
\end{deluxetable}

\clearpage
\begin{deluxetable}{lll}
\tabletypesize{\scriptsize}
\tablecaption{Keplerian Fit to the RV Data for HIP 57050}
\tablewidth{370pt}
\tablehead{Parameter \quad\qquad\qquad\qquad\qquad&  Value (one-planet fit)
\quad\quad\quad\quad\quad&  Value (one-planet fit+trend)\\}
\startdata
$P$ (days)                  & $41.397\pm0.016$  & $41.352\pm0.050$ \\
$m\sin{i}$ ($M_{\rm J}$)\tablenotemark{a} & $0.298\pm0.025$ 
& $0.276\pm0.021$\\
$a$ (AU)\tablenotemark{a}   & $0.163506\pm0.000042$ & $0.16338\pm0.00013$\\
$K$ (m\,s$^{-1}$)           & $37.8\pm4.5$     & $34.0\pm2.8$\\
$e$                         & $0.314\pm0.086$  & $0.194\pm0.073$\\
$\omega$ (deg.)             & $238.1\pm23.2$   & $258.3\pm36.8$\\
MA (deg.)                   & $321.1\pm21.2$   & $273.1\pm55.4$\\
$\chi_{\nu}^2$              & 13.50            & 10.14\\
RMS (m\,s$^{-1}$)           & 9.23             & 8.06\\
trend (m\,s$^{-1}\,d^{-1}$) & -                & $0.00675\pm0.0033$\\
\enddata
\tablenotetext{a}{All elements are defined at epoch 
JD = 2451581.05. 
Uncertainties are based on 1000 bootstrap realizations of
the RV data. We fit a Keplerian orbit to each realization.
The uncertainties are the standard deviations of the
fitted parameters.
Quoted uncertainties in planetary masses and semimajor
axes {\it do not} incorporate the uncertainty in the mass of the star.}
\label{1pl}
\end{deluxetable}

\end{document}